\documentclass[12pt,preprint]{emulateapj}

\usepackage{psfig}
\usepackage{apjfonts}
\usepackage{mathrsfs}

\begin{document}

\title{GALA: an automatic tool for the abundance analysis of stellar spectra\footnotemark[1]}
  
 \footnotetext[1]{Based on observations collected at the ESO-VLT 
 under programs 65.L-0165, 165.L-0263, 073.D-0211, 080.D-0368, 083.D-0208 and 266.D-5655 
 and on data available in the ELODIE archive. This research has also made use of the SIMBAD 
 database, operated at CDS, Strasbourg, France.}

\author{Alessio Mucciarelli$^1$, 
Elena Pancino$^{2,3}$, Loredana Lovisi$^1$, Francesco R. Ferraro$^1$, Emilio Lapenna$^1$}
\affil{$^1$ Dipartimento di Fisica \& Astronomia, Universit\`a 
degli Studi di Bologna, Viale Berti Pichat, 6/2 - 40127
Bologna, ITALY}
\affil{$^2$ INAF - Osservatorio Astronomico di Bologna, Via Ranzani 1 - 40127
Bologna, ITALY}
\affil{$^3$ ASI Science Data Center, I-00044 Frascati, ITALY}

\begin{abstract} 
GALA is a freely distributed Fortran code to derive automatically the atmospheric parameters 
(temperature, gravity, microturbulent velocity and overall metallicity) and 
abundances for individual species
of stellar spectra using the classical method based on the
equivalent widths of metallic lines. 
The abundances of individual spectral lines are derived by using 
the WIDTH9 code developed by 
R. L. Kurucz.
GALA is designed to obtain the best model atmosphere, by optimizing temperature, surface gravity, 
microturbulent velocity and metallicity, after rejecting the discrepant lines. 
Finally, it computes accurate internal errors for each atmospheric parameter and abundance. 
The code permits to obtain chemical abundances and atmospheric parameters for large stellar 
samples in a very short time, thus making GALA an useful tool in the epoch of the multi-object 
spectrographs and large surveys.
An extensive set of tests with both synthetic and observed spectra is 
performed and discussed to explore the capabilities and robustness of the code.

\end{abstract}

\section{Introduction}

The last decade has seen a significant improvement in the study of the chemical composition 
of the stellar populations (in our Galaxy and its satellites), thanks to the 8-10 meters 
class telescopes, coupled with the design of several multi-object mid/high-resolution spectrographs, 
e.g. FLAMES mounted at the Very Large Telescope, 
{AAOmega at the Anglo-Australian Telescope,
DEIMOS at the Keck Observatory and HYDRA at the Blanco Telescope of the 
Cerro Tololo Inter-American Observatory.
These instruments have allowed to enlarge the statistical significance of the acquired stellar spectra, 
but also they have required a relevant effort to manage such large databases. 

The next advent of new surveys aimed to collect huge samples of mid-high 
resolution stellar spectra, as for instance the European Space Agency {\sl Gaia} mission, 
the Gaia-ESO Survey committed by the European Southern Observatory 
\citep{gilmore12}, the 
APOGEE Survey at the Apache Point Observatory \citep{allende08}, 
and the RAVE Survey at the Anglo-Australian Observatory \citep{steinmetz}, 
will make available in real time to the astronomical community an enormous volume of data. 
Other spectroscopic surveys have been already performed, 
like BRAVA \citep{kunder} and ARGOS \citep{freeman13}, both dedicated to the study of the Galactic Bulge.
Also, other multi-object spectrographs are planned or proposed for the next years, 
i.e. HERMES \citep{barden10} at the Anglo-Australian Observatory, 
4MOST \citep{dejong11} at the New Technology Telescope 
and MOONS \citep{cirasuolo11} at the Very Large Telescope.
This perspective, coupled with the huge amount of high quality spectra available in the main 
on-line archives (and not yet totally analysed) clarifies the urgency to develop automatic tools able 
to rapidly and reliably manage such samples of spectra.

%

In the last years several codes aimed at the automatic measurements of the chemical abundances have been 
already developed. They are mainly based on the comparison between the observed spectrum and grids 
of synthetic spectra, for instance ABBO \citep{boni03}, MATISSE \citep{recio06}, SME \citep{valenti}, 
SPADES \citep{posbic}, MyGIsFOS \citep{sbordone10}.
In particular, in these codes the main effort has been devoted to robustly determine the
atmospheric parameters (and hence the elemental abundances) for low ($<$50) 
signal-to-noise (SNR) spectra and generally to develop an algorithm able 
to accurately treat different kind of stars (in terms of metallicity and stellar parameters).

In this paper we present and discuss a new code (named GALA) specifically designed for automatically determining
the atmospheric parameters by using the observed equivalent widths (EWs) of metallic 
lines in stellar spectra, at variance with the majority of the available automated codes. 
GALA is a tool developed within Cosmic-Lab, a 5-years project funded by the European Research Council 
and it is freely available at the website of the project {\sl http://www.cosmic-lab.eu/Cosmic-Lab/Products.html}.

The paper is structured as follows: Section 2 discusses the outline of the classical method 
to derive the main parameters and the interplay occurring among them; Section 3 describes 
the algorithm; Section 4 describes the identification and the rejection of the outliers and Section 5 
discusses other aspects of the code. Section 6 provides a complete description of the 
uncertainties in the calculations. Finally, Sections  7, 8 and 9 discuss a number of tests performed to check 
the stability and the performances of GALA.

\section{The method}

The main advantage of inferring the stellar atmospheric parameters from the EWs is its reproducibility: 
any researcher can directly compare its own results about a given star with other analysis based 
on the same approach.  This allows to distinguish between discrepancies due to the 
method (i.e. the measured EWs) and those due to the physical assumptions of the analysis 
(model atmospheres, atomic data...).
On the other side, one of the most critical aspects of this method is 
the particular accuracy needed in the definition of the linelist, 
by excluding blended lines.
In fact, the codes developed to calculate the abundance from the measured EW
compare the latter with the theoretical strength of the line, changing the abundance 
until the observed and theoretical EW match within a convergence range. 
The theoretical line profile is usually calculated including the continuum opacity sources but 
neglecting the contribution of the neighboring lines \cite[see][for details]{castelli_w}, 
hence the spectral lines to be analysed with this technique are to be checked accurately
against blending (a practice not always performed). Otherwise, the use of synthetic 
spectra allows to also use blended features (and in principle to exploit the information derived 
from all the pixels), but it is more expensive in terms 
of computing time, because large grids of spectra must to be computed, at different 
parameters and with different chemical compositions, and each change in the atomic data leads 
to a recomputation of the synthetic spectra. 

\subsection{The classical spectroscopic method}
\label{fmethod}

The main parameters that define the model atmosphere, 
namely the effective temperature ($T_{\rm eff}$), the surface gravity (log~g), 
the microturbulent velocity ($v_{\rm t}$) and the overall metallicity 
([M/H]\footnote{We adopted the classical {\sl bracket} notation 
where [X/H]=~$A(X)_{star}$-$A(X)_{\odot}$, where A(X)=~$\log{N_{A}/N_{H}}+12$.}) 
are constrained through four constraints: 

\begin{enumerate} 

\item {\sl Temperature:} the best value of $T_{\rm eff}$ is derived by imposing the so-called {\sl excitation equilibrium}, 
requiring that there is no correlation between  the abundance and the excitation potential $\chi$ 
of the neutral iron lines. 
The number of electrons populating each energy level is basically a function of $T_{\rm eff}$, 
according to the Boltzmann equation.
If we assume a wrong $T_{\rm eff}$ in the analysis of a given stellar spectrum, 
we need different abundances for matching the observed profile of transitions with different 
$\chi$. For instance, the use of a value of $T_{\rm eff}$ too large will lead to under-populate the 
lower energy levels, thus the predicted line profile for low-$\chi$ transitions will be 
too shallow and a higher abundance will be needed to match the line profile.
On the other hand, a wrong (too low) $T_{\rm eff}$ will lead to a deeper line profile for the 
low $\chi$ transitions.\\ 
For this reason, a wrong, too large value of $T_{\rm eff}$ will introduce an 
anticorrelation between abundances and $\chi$, and in the same way, 
a positive correlation is expected in the case of the adoption of a value of 
$T_{\rm eff}$ too small;\\

\item {\sl Surface gravity:} the best value of log~g is derived with the so-called 
{\sl ionization equilibrium} method, requiring that for a given species, the same abundance 
(within the uncertainties) has been obtained from lines of two 
ionization states (typically, neutral and singly ionized lines). 
Being the gravity a direct measure of the pressure of the photosphere, variations of log~g 
lead to variations of the ionized lines (very sensitive to the electronic pressure), while the neutral 
lines are basically insensitive to this parameter.
This method assumes implicitly 
that the energy levels of a given species are populated according 
to the Boltzmann and Saha equations (thus, under Local Thermodynamical Equilibrium (LTE) conditions). 
Possible departures from this assumption (especially critical for metal-poor 
and/or low-gravity stars) could alter the derived gravity when it is derived from 
the ionization balance, because non-LTE effects affect mainly the neutral lines 
(even if the precise magnitude of the departures from LTE for the iron lines is still matter of debate).
As sanity check, following the suggestion by \citet{edv88}, surface gravities determined from the 
ionization equilibria have to be {\sl "checked - when possible - with gravities determined 
from the wings of pressure broadened metal lines"};\\

\item {\sl Microturbulent velocity:} $v_{\rm t}$ is computed by requiring that there is no correlation 
between the iron abundance and the line strength 
\citep[see][for a discussion about different approaches]{m11rn}.
The microturbulent velocity affects mainly the moderate/strong lines located along 
the flat regime of the curve of growth, while 
the lines along the linear part of the curve of growth are mainly sensitive to the abundance 
instead of the velocity fields.
The necessity to introduce the microturbulent velocity as an additional broadening 
(added in quadrature to the Doppler broadening)
arises from the fact that the non-thermal motions (basically due to the onset of the convection 
in the photosphere) are generally not well described by the 1-dimensional, static model atmospheres.
Citing \citet{kurucz05}, {\sl microturbulent velocity is a parameter that is generally 
not considered physically except in the Sun,}  because in the Sun the velocity fields can be derived 
as a function of the optical depth through the analysis of the intensity spectrum 
\citep[as performed by][]{fontenla}. For the other stars 
$v_{\rm t}$ represents only a corrective factor that minimizes the line-to-line scatter for a given species 
and it compensates (at least partially) to the incomplete description of the convection as implemented 
in the 1-dimensional model atmospheres;\\
 
\item {\sl Metallicity:} [M/H] is chosen according to the average iron content 
of the star, assuming [Fe/H] as a proxy of the overall metallicity. 
Generally, [Fe/H] is adopted as a good proxy of the metallicity because of its 
large number of available lines, but it does not indicate necessarily the overall metallicity 
of the studied star. In fact, iron is generally not the most abundant element in the stars, while 
elements as C, N and O would be the best tracers of the stellar metallicity (but they are difficult to measure). 

\end{enumerate}

Because of its statistical nature,
the spectroscopic optimization of all the parameters simultaneously can be performed only 
if we have a sufficient number of Fe lines, distributed in a large range 
of EW and $\chi$ and in two levels of ionization. Alternatively, $T_{\rm eff}$ and log~g can be 
inferred from the photometry (for instance with the isochrone-fitting technique or employing 
empirical or theoretical $T_{\rm eff}$--color relations) 
or by fitting the wings of damped lines (as the hydrogen Balmer lines or the Mg {\sl b} triplet) 
sensitive to the parameters, 
and only $v_{\rm t}$ needs to be tuned spectroscopically (following the approach described above). 
Note that some authors consider the method to derive the parameters based on
these constraints only as sanity checks performed 
{\sl a posteriori} on the photometric parameters, while other authors rely on these 
constraints to infer the best parameters.

\subsection{The interplay among the parameters}
\label{interplay}

In light of the method described above, it is worth to bear in mind that the atmospheric 
parameters are correlated with each other.
In fact, the  strongest lines are typically 
those with low $\chi$: Fig.~\ref{kur_ep} plots all the transitions 
in the range $\lambda$=~4000-8000 $\mathring{A}$ and with $\chi<$10 eV in the Kurucz/Castelli 
database\footnote{http://wwwuser.oat.ts.astro.it/castelli/linelists.html}
in the plane $\chi$ vs $\log(gf)-\theta\chi$ 
\footnote{The term $\log(gf)-\theta\chi$ is used as theoretical 
proxy of the line strength, where log(gf) is the oscillator strength and 
$\theta$=~5040/$T_{\rm eff}$ $eV^{-1}$.}. 
As mentioned in Section~\ref{fmethod}, $T_{\rm eff}$ and $\chi$ are strictly linked, 
and there is also a connection between $v_{\rm t}$ and the line strength. Hence,  
the statistical correlation between $\chi$ and the 
line strength leads to a correlation between $T_{\rm eff}$ and $v_{\rm t}$. Thus, 
a variation of $T_{\rm eff}$ implies a variation of $v_{\rm t}$. Also, variations of $T_{\rm eff}$ 
and $v_{\rm t}$ will change differently the abundances derived from different 
levels of ionization (hence, the gravity).

\begin{figure}[h]
\epsscale{1.2}   
\plotone{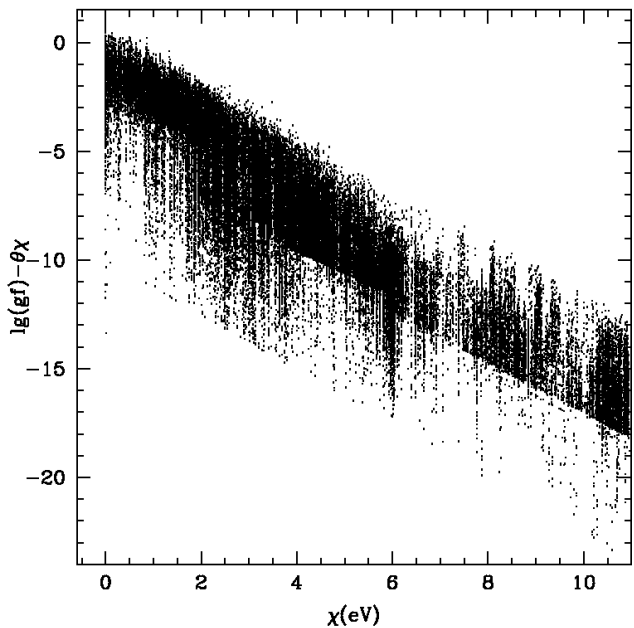}
\caption{Behaviour of the line strength (computed assuming $T_{\rm eff}$=4500 K) as 
a function of the excitation potential $\chi$ for all the transitions available in the Kurucz/Castelli linelist 
in the range of wavelength $\lambda$=4000-8000 $\mathring{A}$ and with $\chi<$10 eV. }
\label{kur_ep}
\end{figure}

Let us consider an ATLAS9 model atmosphere computed with $T_{\rm eff}$=4500 K, log~g=1.5, $v_{\rm t}$=2 km/s 
and [M/H]=--1.0 dex \citep{castelli04} and a set of neutral and singly ionized iron lines (predicted to be unblended 
through the inspection of a synthetic spectrum calculated with the same parameters and convoluted 
at a spectral resolution of 45000).
The EWs of these transitions are computed by integrating the theoretical line profile 
through the WID subroutine implemented in the WIDTH9 code \citep{castelli_w}.
This means that each of these EWs will provide exactly [El/H]=--1.0 dex when they are analysed 
by using the model atmosphere described above.

The analysis of these lines (adopting always the same set of EWs) is repeated investigating 
a regular grid of the atmospheric parameters, 
namely $T_{\rm eff}$=~3600--5400 K, log~g=~0.5--2.5, $v_{t}$=~1.0--3.0 km/s, steps 
of $\delta T_{\rm eff}$=~200 K, $\delta logg$=~0.2, $\delta v_{t}$=~0.5 km/s and assuming 
for all the models [M/H]=~--1.0 dex.
Fig.~\ref{int_t}, \ref{int_v} and \ref{int_g} display the quite complex interplay 
occurring among the atmospheric parameters.

\begin{itemize} 

\item Fig.~\ref{int_t} shows the behaviour of the slope $S_{\chi}$ of the 
A(Fe)--$\chi$ 
relation for the above sample of lines as a function of $T_{\rm eff}$, keeping gravity fixed, 
but varying the microturbulent velocity. The thick grey line connects 
points calculated with the original $v_{\rm t}$ of the model. The global trend 
is basically linear, at least if we consider a range of $\pm$1000 K around the 
original $T_{\rm eff}$.
The inset panel shows the behaviour of $T_{\rm eff}$ for which $S_{\chi}$  is zero (thus, the 
best $T_{\rm eff}$) as a function of $v_{\rm t}$ and considering different gravities. 
The derived best temperature increases with increasing $v_{\rm t}$ (at fixed log~g); 
the gravity has only a second-order effect and it does not change the general behaviour 
of the best $T_{\rm eff}$ as function of $v_{\rm t}$.\\

\begin{figure}[h]
\epsscale{1.2} 
\plotone{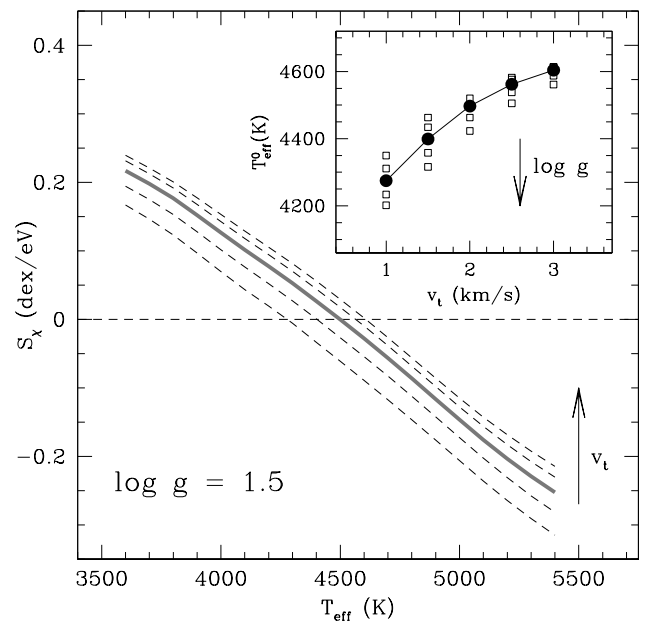}
\caption{Main panel: behaviour of the slope $S_{\chi}$ of the A(Fe)--$\chi$ relation as a function 
of $T_{\rm eff}$ assuming log~g=~1.5 and for different values of $v_{t}$ (dashed curves). The 
thick grey curve represents the behaviour computed for the original microturbulent velocity 
($v_{t}$=~2.0 km/s).
Horizontal dashed line is the zero value (according to the excitation equilibrium). 
The inset panel shows the behaviour of 
the best value of $T_{\rm eff}$ (for which $S_{\chi}$=~0) as a function of the used $v_{\rm t}$ (empty squares) 
and for different gravities; black points are referred to the original gravity (log~g=1.5).
}
\label{int_t}
\end{figure}

\item Fig.~\ref{int_v} summaryzes the behaviour of the slope $S_{EWR}$ of the A(Fe)--EWR relation 
as a function of $v_{\rm t}$ 
(where EWR indicates the reduced EW, defined as EWR=$\lg(EW/\lambda)$), 
keeping gravity fixed at the original value but varying 
$T_{\rm eff}$. For a given temperature, the slope decreases increasing $v_{\rm t}$, with 
a behaviour that becomes less steep at $v_{\rm t}$ larger than the original value.
The effect of the $T_{\rm eff}$ is appreciable for $T_{\rm eff}$ larger than the original 
value, while for lower $T_{\rm eff}$ all the curve are very similar to each other.
The inset shows the change of the best value of $v_{\rm t}$ (for which the slope of the A(Fe)--EWR relation is zero)
as a function of $T_{\rm eff}$ and for different log~g. The observed trend is quite complex: 
basically, we note that the best value of $v_{\rm t}$ is very sensitive to $T_{\rm eff}$, when 
the latter is overestimated with respect to the true temperature, but with a 
negligible dependence from gravity, while the behaviour is the opposite when 
$T_{\rm eff}$ is under-estimated, with a degeneracy between $T_{\rm eff}$ and the best 
$v_{t}$ but a consistent dependence from gravity.\\

\begin{figure}[h]
\epsscale{1.2} 
\plotone{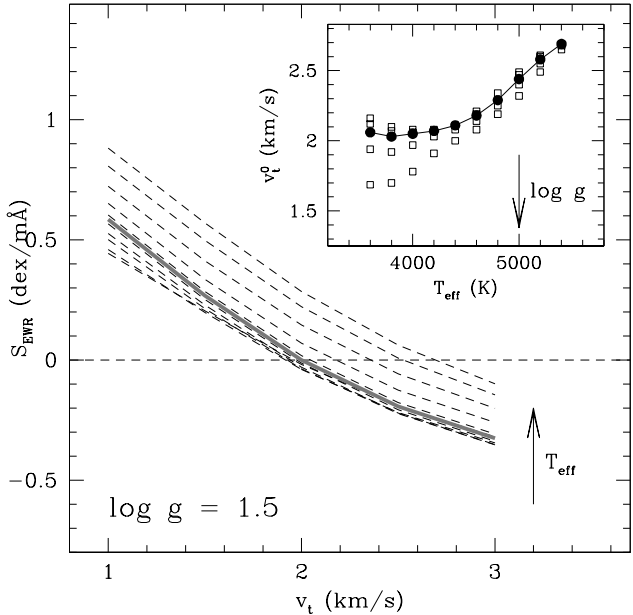}
\caption{Main panel: behaviour of the slope $S_{EWR}$ of the  A(Fe)--EWR  relation as a function 
of $v_{t}$ assuming log~g=~1.5 and for different $T_{\rm eff}$ (dashed curves). The 
thick grey curve represents the behaviour computed for the original temperature ($T_{\rm eff}$=~4500 K).
Horizontal dashed line is the zero value. The inset panel shows the behaviour of 
the best value of the microturbulent velocity $v_{t}$ 
(for which $S_{EWR}$=~0) as a function of the used $T_{\rm eff}$ (empty squares) 
and for different values of gravities; black points are referred to the original gravity (log~g=1.5).}
\label{int_v}
\end{figure}

\item Fig.~\ref{int_g} shows the behaviour of the difference between A(Fe~I) and A(Fe~II) 
as a function of log~g, assuming $v_{\rm t}$=~2 km/s and for different $T_{\rm eff}$. The general behaviour 
is linear and the iron difference increases considerably increasing temperature.
The best value of gravity (see the inset in Fig.~\ref{int_g}) is highly sensitive to changes 
in $T_{\rm eff}$, with $\delta$logg/$\delta T_{\rm eff}\simeq$1 dex/300 K in the investigated case, 
but for a fixed $T_{\rm eff}$ turns out to be marginally sensitive to $v_{\rm t}$ 
(this is due to the fact that the Fe~II lines are basically distributed in strength 
in a similar way to the Fe~I lines).

\begin{figure}[h]
\epsscale{1.2} 
\plotone{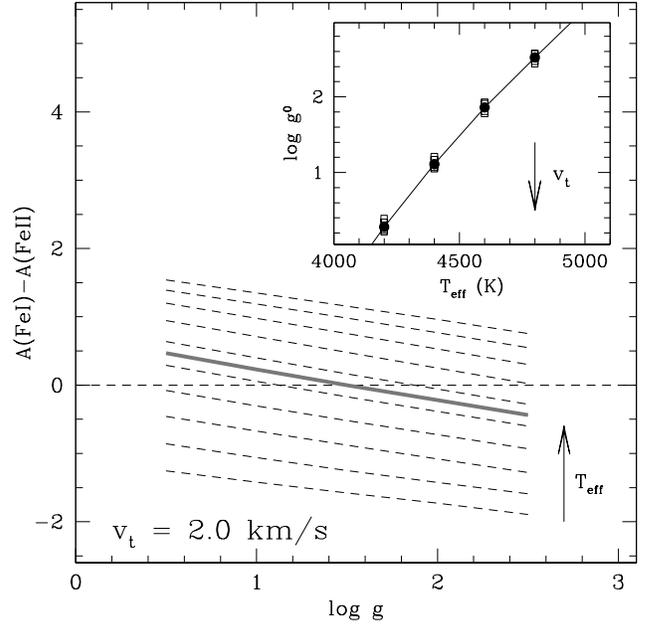}
\caption{Main panel: behaviour of the difference between A(Fe~I) and A(Fe~II) as a function 
of log~g assuming $v_{\rm t}$=~2.0 km/s and for different $T_{\rm eff}$ (dashed curves). The 
thick grey curve represents the behaviour computed for the original temperature ($T_{\rm eff}$=~4500 K).
Horizontal dashed line is the zero value (according to the ionization equilibrium). 
The inset panel shows the behaviour of 
the best log~g (for which A(Fe~I)=~(Fe~II)) as a function of the used $T_{\rm eff}$ (empty squares) 
and for different gravities; black points are referred to the original microturbulent velocity
($v_{\rm t}$=~2.0 km/s).}
\label{int_g}
\end{figure}

\end{itemize}

It is important to bear in mind that the these considerations are appropriate for the investigated 
case of a late-type star but the dependencies among the parameters can be different for different 
regimes of atmospheric parameters and/or metallicity.
However, the example presented above demonstrates that an analytic approach to derive the best model atmosphere 
is discouraged because it needs to know the precise topography of the parameters space and requires 
the inspection of a large number of model atmospheres.

\section{GALA}

GALA is a program written in standard Fortran 77, that uses the WIDTH9 code 
developed by R. L. Kurucz in its Linux version \citep{sbordone04} to derive 
the chemical abundances of single, unblended absorption lines starting from 
their measured EWs. 
We used our own version of WIDTH9, modified in order to have a more flexible format for 
the input/output files with respect to the standard version of the code available 
in the website of F. Castelli, while the input physics and the method to derive 
the abundances are unchanged.

GALA is specifically designed to 
\begin{enumerate}
\item choose the best model atmosphere by using the observed EWs 
of metallic lines;
\item manage the input/output files of the WIDTH9 code;
\item provide statistical and graphical tools to evaluate the quality of the 
final solution and the uncertainty of the derived parameters. 
\end{enumerate}

GALA is designed to handle both ATLAS9 
\citep{kur04} and MARCS \citep{gustaf98} model atmospheres, that are the most 
popular employed dataset of models. 
The current version has been compiled with the Intel Fortran Compiler (versions 11, 12 and 13) 
and tested on the Leopard, Snow Leopard and Lion Mac OSX systems, and on the Ubuntu,
Fedora and Mandriva Linux platforms.

\subsection{Optimization parameters}
\label{cxopt}

GALA has been designed to perform the classical chemical analysis based on the EWs 
in an automated way.
The user can choose to perform a full spectroscopic optimization of the parameters or 
to optimize only some parameters, keeping the other parameters fixed to the specified input values.

The algorithm optmizes one parameter at a time, checking continuously if the new value 
of a given parameter changes the validity of the previously optmized ones. 

For each atmospheric parameter {\sl X} (corresponding to $T_{\rm eff}$, log~g, $v_{\rm t}$ and [M/H]) 
we adopt a specific {\sl optimization parameter C(X)}, defined in a way that it turns out to be 
zero when the best value of the {\sl X} parameter has been found. 
Hence, GALA varies {\sl X} until a positive/negative pair of the {\sl C(X)} is found, thus bracketing the zero value 
corresponding to the best value.
Thus, the condition {\sl C($\tilde{\rm X}$)}=~0 identifies X=$\tilde{\rm X}$ 
as best value of the given parameter.
Finally, the best solution converges to a set of parameters which verifies 
simultaneously the constraints described in Section \ref{fmethod}.
When the atmospheric parameters have been found, the abundances of all 
the elements for which EWs have been provided are derived.


According to the literature,
the adopted {\sl C(X)} have been defined to parametrize the conditions listed in 
Section \ref{fmethod}: 
\begin{enumerate}
\item the angular coefficient of the A(Fe)--$\chi$ relation ($S_{\chi}$)
to constrain $T_{\rm eff}$; the lower panel of Fig.~\ref{slope} shows the change 
of this slope for a set of theoretical EWs computed with the correct $T_{\rm eff}$ 
(grey points) and with temperatures varied by $\pm$500 K (empty points).
The variation of $T_{\rm eff}$ produces a change of the slope (but also a 
change in the y-intercept);
\item the angular coefficient of the A(Fe)--EWR relation ($S_{EWR}$)
to constrain $v_{\rm t}$. 
The upper panel of Fig.~\ref{slope} shows the same set of theoretical EWs 
analysed with different $v_{\rm t}$;
\item the difference of the mean abundances obtained from Fe~I and Fe~II lines to constrain 
the gravity;
\item the average Fe abundance to constrain the metallicity of the model.
\end{enumerate}

\begin{figure}[h]
\epsscale{1.2} 
\plotone{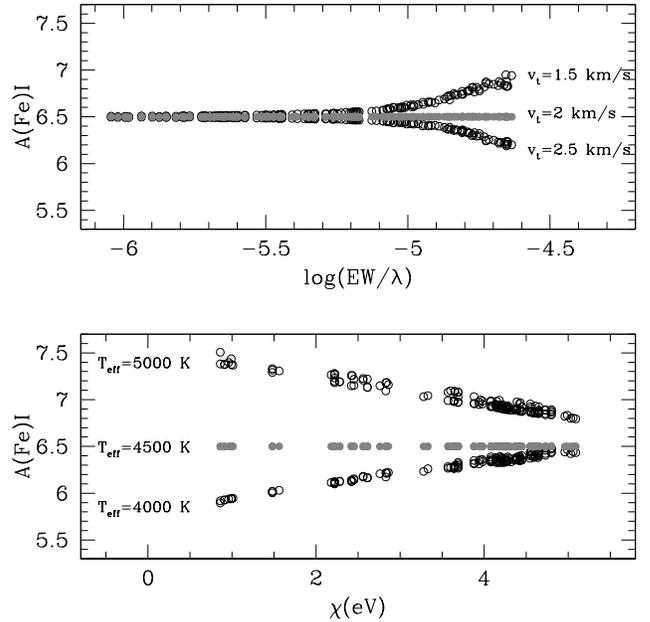}
\caption{Upper panel: behaviour of A(Fe~I) as a function of the reduced 
equivalent widths for a set of theoretical EWs obtained from a model atmosphere 
computed with $T_{\rm eff}$=~4500 K, log~g=~2.0 and $v_{\rm t}$=~2 km/s. The derived 
abundances are obtained by adopting the correct value of $v_{\rm t}$ (grey points) and 
two {\sl wrong} values of $v_{\rm t}$ (open circles). 
Lower panel: behaviour of A(Fe~I) as a function of the excitation potential 
for the same set of theoretical EWs. Grey points are the results obtained 
by analyzing the lines with the correct value of $T_{\rm eff}$, while the empty circles 
are obtained by over/under-estimate $T_{\rm eff}$ by $\pm$500 K.}
\label{slope}
\end{figure}

If the errors in the EW measurement ($\sigma_{EW}$) are provided as an input, the slopes are computed 
taking into account the abundance uncertainties of the individual lines; 
the uncertainty on the iron abundance of a given line is estimated from the 
difference of the iron abundance computed for the input EW and for EW+$\sigma_{EW}$
\footnote{The uncertainties in A(Fe) are assumed symmetric with respect to the variations 
of the EW ($\pm\sigma_{EW}$); we checked that this assumption is correct at a level of $\sim$0.01 dex.}.
In the A(Fe)--$\chi$ plane the uncertainties of $\chi$ are reasonably assumed negligible, because 
the uncertainties of $\chi$ are typically less than 0.01 eV, while 
in the A(Fe)--EWR plane the least square fit takes into account the uncertainties in both the 
axis \citep[following the prescriptions by][]{press}.

The flexibility of GALA permits to simultaneously deal with stars of different spectral types. 
This is made by considering that a large number of Fe~I lines is generally available for F-G-K spectral types stars, 
whereas they are less numerous (or lacking) in O-B-A stars, for which a large number of Fe~II lines 
is typically available. Also, in some spectral regions there is a large number of lines for other 
iron-peak elements (mainly Ni, Cr and Ti). For this reason, 
GALA is designed to optimize the parameters also using other lines instead of Fe~I lines, by  
appropriately configuring the code. In the following, we will refer to the optimization 
made by using Fe~I for $T_{\rm eff}$ and $v_{\rm t}$, but our considerations are valid also for other 
elements with a sufficient number of lines.


\subsection{The main structure}

GALA is structured in three main Working-Blocks:\\ 
(1)~the {\sl Guess Working-Block} is aimed at finding the guess atmospheric parameters 
in a fast way (this is especially useful in cases of large uncertainties 
or in lacking of first-guess value for the parameters);\\ 
(2)~the {\sl Analysis Working-Block} which finds the best model atmosphere through a 
local minimization and starting from the guess parameters provided 
by the user or obtained through the previous block;\\ 
(3)~the {\sl Refinement Working-Block} that refines the solution, starting from the 
atmospheric parameters obtained in the previous block.\\

GALA can be flexibly configured to use different combinations of the three main 
Working-Blocks. We defer the reader to Section~\ref{stabi} for a discussion of the 
effects of the Working-Blocks.
In the following we describe 
the algorithm of each block and the cases in which they are recommended.

\subsubsection{Guess Working-Block}

If the atmospheric parameters are poorly known, this Working-Block verifies 
them quickly by exploring the parameters space in a coarse grid. Thus, it saves 
a large amount of time if the initial parameters are far away from the correct solution.

\begin{enumerate}
\item
As a first step, the abundances for each line are derived with the input parameters and 
some lines are labelled as outliers and excluded from the analysis 
(the criteria of the rejection are described in Sect.~\ref{weed}).
The surviving lines will be used in this Working-Block and no other line rejection 
will be performed until convergence.
\item
The metallicity of the model is eventually readjusted according to the average 
iron abundance.
\item
$S_{\chi}$ is computed with the input $T_{\rm eff}$ and with a $T_{\rm eff}$ varied by +500K 
(if $S_{\chi}$ is positive) or --500 K (if $S_{\chi}$ is negative). 
This procedure is repeated until a pair of positive/negative values of $S_{\chi}$ is found, thus 
to bracket the $T_{\rm eff}$ value for which $S_{\chi}$=~0. 
The behaviour of $S_{\chi}$ as a function of $T_{\rm eff}$ is described with a linear relation, 
finding the value of $T_{\rm eff}$ for which $S_{\chi}$=~0. The description of this relation with 
a linear fit is legitimate  as long as the employed $T_{\rm eff}$ range is relatively small (in this case 
500~K), because for larger range the 
behaviour of $S_{\chi}$ as a function of $T_{\rm eff}$ could become non linear (mainly due 
to the interplay with the other parameters).
\item
The new value of $T_{\rm eff}$ is adopted to find a new value of $v_{\rm t}$, following the same approach 
used for $T_{\rm eff}$ and searching for a positive/negative pair of $S_{EWR}$ over a range 
of 0.5 km/s.
\item
Finally, a new value of log~g is found, starting from the $T_{\rm eff}$ and $v_{\rm t}$ derived 
above, by searching for a positive/negative pair of $\Delta(Fe)$, over a range of 0.5 dex 
in gravity. 
\end{enumerate}

The entire procedure (from the optimization of [M/H] to that of log~g) is repeated 
for a number of iterations chosen by the user and finally a new set of input parameters is found. 
Generally three/four 
iterations are sufficient to find a good solution.
The final solution is accurate enough to identify the neighborhood of the real solution 
in the parameters space but it could be unreliable 
since it needs to be checked for the covariances among the parameters (which is 
the task of the {\sl Analysis Working-Block}.

\subsubsection{Analysis Working-Block}

This Working-Block performs a complete optimization 
starting from the input parameters provided by the users or from those 
obtained with the {\sl Guess  Working-Block}.  This block is developed to find a robust 
solution under the assumption that the input values are reasonably close to 
the real solution. When good priors are available (for instance, in the case of 
stellar cluster stars) this block is sufficient to find the solution, without 
the use of the guess block. Otherwise, when the guess model is uncertain 
(for instance, in the case of field stars for whose reddening, distance and evaluative 
mass could be highly uncertain, or, generally, in the case of inaccurate photometry), 
the analysis block is recommended to be used after the guess block. 
Fig.~\ref{diagram} shows as a flow chart the main steps of this Working-Block. 

\begin{figure}[h]
\epsscale{1.2} 
\plotone{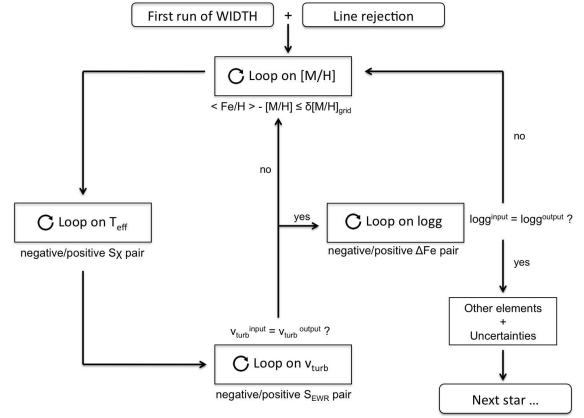}
\caption{Flow diagram for the {\sl Analysis Working-Block} of GALA (see Section 3.2 for details).}
\label{diagram}
\end{figure}

The following iterative procedure is performed:
\begin{enumerate} 
\item 
The procedure starts by computing the abundances using the guess parameters and 
performing a new line rejection (independent from that of the previous block).
At variance with the previous Working-Block, now the parameters are varied by little steps, 
configured by the user. 
\item
The model metallicity is refined to match the average iron abundance.
\item
A new model with different $T_{\rm eff}$ is computed, according to the sign of 
$S_{\chi}$ of the previous model (i.e., a negative slope indicates a overestimated $T_{\rm eff}$ 
and vice versa). New models, varying only $T_{\rm eff}$, are computed until a 
pair of negative/positive $S_{\chi}$ is found.  Thus, these two values of $T_{\rm eff}$  
identify the range of $T_{\rm eff}$ where the slope is zero.
$T_{\rm eff}$ corresponding to the minimum $|S_{\chi}|$ is adopted. 
\item 
The same procedure is performed for $v_{\rm t}$. If the final value of $v_{\rm t}$
is different from that used in the previous loop, GALA goes back to item (2), checking if the 
new value of $v_{\rm t}$ needs a change in [M/H] and $T_{\rm eff}$. Otherwise, the procedure moves on to the 
next loop.
\item
The surface gravity is varied until a positive/negative pairs of $\Delta(Fe)$ is found. 
If the output log~g differs from the input value, GALA returns to (2) with 
the last obtained model atmosphere and the entire procedure is repeated.
\item When a model that satisfies simultaneously the four constraints is found, the 
procedure ends and the next star is analysed.
\end{enumerate}

We stress that the method employed in this Working-Block is very robust but it has the
disadvantage of being slow if the guess parameters are far from the local solution.

\subsubsection{Refinement Working-Block}
This block allows to repeat the previous Working-Block 
using the solution obtained in the previous block as a starting point. 
A new rejection of the outliers is performed and the same approach of the 
{\sl Analysis Working-Block} is used. This block can be useful 
to refine locally the solution when the first block is switched off.\\ 

The main advantage of the {\sl Refinement Working-Block} is that the new line-rejection is performed 
by using accurate atmospheric parameters (because obtained from the {\sl Analysis Working-Block}). 
As will be discuss in Section \ref{weed}, the line-rejection performed on abundance distributions 
obtained with wrong parameters can be risky, losing some useful lines.

\section{Weeding out the outliers}
\label{weed}

The detection and the rejection of lines with discrepant abundances are crucial aspects 
of the procedure and require some 
additional discussion. Before ruling out a line from the line list, we need to 
understand the origin of the detected discrepancy. Basically, the main reasons for a discrepant 
abundance are:\\ 
(1)~inaccurate atomic data (i.e. oscillator strengths) that can under- or over-estimate 
the abundance;\\ 
(2)~unrecognized blends with other lines (providing 
systematically overestimated abundances);\\ 
(3)~inaccurate EW measurement.\\ 
The first two cases can be partially avoided with an effort during 
the definition of the adopted linelist, including only transitions 
with accurate log~gf and checking each transition against blending, 
according to the atmospheric parameters and the spectral resolution. 

GALA rejects the lines according to the following criteria:
\begin{enumerate}

\item lines weaker or stronger than the input EWR thresholds are rejected, in order to exclude 
either weak and/or strong lines. 
In fact, weak lines can be heavily affected by the noise, whereas strong lines can be 
too sensitive to $v_{\rm t}$ and/or they can have damping wings for which the fit with a 
Gaussian profile could be inappropriate, providing a systematic under-estimate of the EW;

\item lines whose uncertainty on the EW measurement 
is larger than an input threshold chosen by the user and expressed as a percentage.
Note that not all the codes developed to measure EWs provide an estimate of the EW error, 
despite the importance of this quantity. 
For instance, among the publicly available codes aimed to measure EWs, DAOSPEC \citep{stetson} 
and EWDET \citep{ramirez01} provide accurate uncertainty evaluations for each line, while 
SPECTRE \citep{sneden} and ARES \citep{sousa07} do not include EW error calculations.
For this reason GALA works even if $\sigma_{EW}$
are not provided, although this affects the final solution accuracy, because all the transitions 
will be weighed equally, despite their different measurement quality;

\item 
lines are rejected according to their distance from the 
best-fit lines computed in the A(Fe)--$\chi$ and A(Fe)--EWR planes 
through a $\sigma$-rejection algorithm.
A $\sigma$-rejection from the best-fit lines in the planes used for the 
optimization is more robust with respect to a simple $\sigma$-rejection based on the abundance 
distribution. In the latter case, there is the risk to lose some lines important for the 
analysis, thus biassing the results. Fig~\ref{outlier} explains this aspect: we consider a synthetic spectrum of a 
giant star ($T_{\rm eff}$=~4500 K) and we measure the EWs after the injection of Poissonian noise 
in the spectrum in order to reproduce a reasonable good SNR ($\sim$30). 
Fig.~\ref{outlier} shows the distribution of the 
Fe~I lines in the A(Fe)--$\chi$ plane when the chemical analysis is performed by using a wrong model 
atmosphere with $T_{\rm eff}$=~5200 K (thus leading to an anticorrelation between A(Fe~I) and $\chi$). 
In the upper panel the outliers were rejected according to 
the median value of the abundance distribution, shown as gray solid line, while the two dashed lines 
mark $\pm{\rm 3}\sigma$ level and black points are the surviving lines. 
In the lower panel the 
rejection is performed according to the distance from the best-fit line 
(shown as solid line while the two dashed lines mark $\pm{\rm 3}\sigma$ level). It is evident that 
in the first case the majority of the discarded lines are those with low $\chi$ (thus, the 
most sensitive to the $T_{\rm eff}$ changes), with the risk of introducing a bias in the $T_{\rm eff}$ 
determination. On the other hand, the method of rejection shown in the lower panel of 
Fig.~\ref{outlier} preserves the low-$\chi$ lines, guaranteeing the correctness of the final 
solution.
\end{enumerate}

\begin{figure}[h]
\epsscale{1.2} 
\plotone{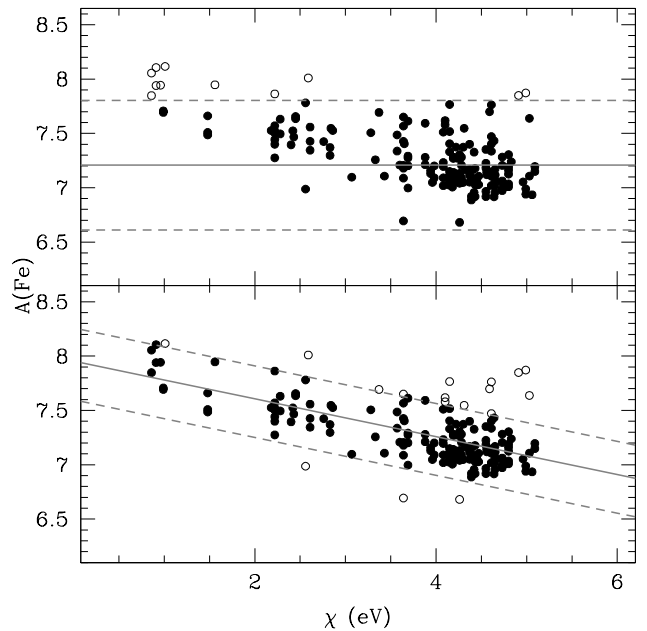}
\caption{Behaviour of the Fe~I abundances as a function of the excitation potential 
for a synthetic spectrum computed assuming $T_{\rm eff}$=~4500 K but analysed with a model 
atmosphere with $T_{\rm eff}$=~5200 K. Black circles are the lines survived after the 
line-rejection procedure and the empty points the rejected lines. Upper panel shows 
the results by adopting a rejection based on the abundance distribution; the solid line 
indicates the median abundance and the dashed lines mark $\pm{\rm 3}\sigma$ level.
The lower panel shows the results by using the procedure employed by GALA: solid line is 
the best-fit line and the dashed lines mark $\pm{\rm 3}\sigma$ from the best-fit line.}
\label{outlier}
\end{figure}

An important point is that the outlier rejection in GALA is not performed 
independently in each iteration of the code, but only at the beginning of each 
Working-Block. This is especially important, because it allows
to use always the same sample of lines during the optimization process, avoiding the risk to introduce 
spurious trends in the 
behaviour of the given {\sl optimization parameter} as a function of the corresponding 
atmospheric parameters. 
In fact, the values of C(X) derived from two different sets of lines of the same spectrum, but for 
which an independent rejection of the outliers has been performed, cannot be directly 
compared to each other to derive X.
In particular, this effect is magnified in cases of small number of lines, where the 
impact of the lines rejection can be critical.

\section{More details}

\subsection{A comment about the gravity}
The most difficult parameter to be constrained with the classical 
spectroscopic method is the gravity. 
This because of the relatively small number of available Fe~II lines, which can vary 
in the visual range from a handful of transitions up to $\sim$20, depending 
on the spectral region and/or the metallicity (for instance, some 
high-resolution spectra with a small wavelength coverage, as the GIRAFFE@VLT or 
the Hydra@BlancoTelescope spectra, can be totally lacking in Fe~II lines). 

GALA is equipped with different options to optimize log~g:\\ 
(1)~the normal optimization by using the difference between the average 
abundances from neutral and singly ionized iron lines (as described above);\\ 
(2)~the gravity is computed from the Stefan-Boltzmann equation, by providing in 
input the term $\epsilon={\rm log(4GM\pi\sigma/L)}$, where G is the gravitational constant, 
$\sigma$ the Boltzmann constant and M and L are the mass and the luminosity of the star. 
Thus, during the optimization process, the gravity is re-computed 
(as log~g=~$\epsilon$-4$\log{T_{\rm eff}}$) in each iteration 
according to the new value of $T_{\rm eff}$;\\ 
(3)~the gravity is computed by assuming a quadratic relation 
log~g=~A+B$\cdot T_{\rm eff}$+C$\cdot T_{\rm eff}^2$ and providing 
in input the coefficients A, B, C. This option is useful when the investigated stars 
belong for instance to the same stellar cluster and log~g and $T_{\rm eff}$ can be parametrized 
by a simple relation (i.e. as that described by a theoretical isochrone for a given evolutionary stage). 

The user can choose the way to treat log~g (fixed or optimized following one of the 
methods described above); if the optimization of log~g from the iron lines is requested 
but no Fe~II lines are available, GALA will try to use the second option (lines of other elements 
in different stages of ionization), or eventually will fix log~g to the input value.

\subsection{Model atmospheres}

The algorithm used in GALA is basically independent from the code adopted
to derive the abundances and from the model atmospheres. 
GALA is designed to manage both the two most used and publicly available models atmospheres, 
namely ATLAS9 and MARCS:\\ 
\begin{itemize}
\item{\bf ATLAS9}: 
The suite of Kurucz codes represents the only suite of open-source and free programs to face 
the different aspects of the chemical analysis (model atmospheres, abundance calculations, 
spectral synthesis), allowing any user to compute new models and upgrade  parts of the codes. 
GALA includes a dynamic call to the ATLAS9 code
\footnote{The ATLAS9 source code is available at the website http://wwwuser.oat.ts.astro.it/castelli/sources/atlas9codes.html}.
Any time GALA needs to investigate a given set of atmospheric parameters, ATLAS9 is called, a new 
model atmosphere is computed and finally it is 
stored in a directory. The latter is checked by GALA whenever a model atmosphere is requested, 
and ATLAS9 called only if the model is lacking. 
The convergence of the new model atmosphere is checked for each atmospheric layer. Following the 
prescriptions by \citet{castelli88}, we require errors less than 1\% and 10\% for the flux and 
the flux derivative, respectively. 
Additional information about the 
calculation for each model atmosphere is saved.
In the current version, GALA is able to manage the grid of ATLAS9 models by \citet{castelli04} 
and the new grid of models calculated by \citet{meszaros} for the APOGEE survey.
\item{\bf MARCS}:
At variance with ATLAS9, for the MARCS models the code to compute new model atmospheres 
is not released to the community. However, the Uppsala group provides a large grid of the 
MARCS models on their website\footnote{http://marcs.astro.uu.se/}. When GALA works with these 
grids (including both plane-parallel and spherical symmetry), new models are computed by interpolating 
into the Uppsala grid by using the code developed by T. Masseron \citep{masse}
\footnote{The original code is available at http://marcs.astro.uu.se/software.php.}.
This code has been modified in 
order to put the interpolated MARCS models in ATLAS9 format to use with WIDTH9.

\end{itemize}

Note that the automatization of the chemical analysis based on EWs needs
a wide grid of model atmospheres (both to interpolate and compute new models) 
linked to the code in order to freely explore the parameter space.
Thus, GALA is linked to the ATLAS9 grid both with solar-scaled and $\alpha$-enhanced 
chemical composition and to the MARCS grid with standard composition. 
Also, the use of other models or model grids can be easily implemented in the code.
Sometimes, peculiar analysis or tests need to use specific models, 
for instance for the Sun 
(see the set of solar model atmospheres available in the 
website of F. Castelli)
or with arbitrary chemical compositions, as those computed 
with the ATLAS12 code \citep{castelli_12}.
When a specific, single model is called, all the optimization options are automatically swichted off.

\subsection{Exit options}
GALA is equipped with a number of exit flags in order to avoid infinite loops or
unforeseen cases stopping the analysis of the entire 
input list of stars. We summarize here the main exit options:\\ 
(1)~the user can set among the input parameters the maximum number 
of allowed iterations for each star. When the code reaches this value 
it stops the analysis, moving to the next star. Generally this parameter 
depends on the adopted grid steps and if the input atmospheric parameters are close 
or not to the real parameters. When the {\sl Guess Working-Block} is used, typically the 
analysis block converges in 3-5 iterations;\\ 
(2)~if the dispersion around the mean of the abundances of the lines used for the optimization 
(after the line rejection) exceeds 
a threshold value GALA skips the star. In fact, 
very large dispersions can possibly suggests some problems in the EW measurements;\\ 
(3)~if the number of the lines used for the optimization (after the line rejection) 
is smaller than a threshold, the optimization 
is not performed and the atmospheric parameters are fixed to the input guess values;\\ 
(4)~the procedure is stopped if the requested atmospheric parameter is outside 
the adopted grid of model atmospheres;\\ 
(5)~GALA skips the analysis of the star if the call to the model atmosphere fails 
(problems in the ATLAS9 models computation or in the MARCS models interpolation) 
and the required model is not created. Otherwise, if the ATLAS9 model is calculated 
but some atmospheric layers do not converge (according to the 
criteria discussed above) GALA continues the analysis but it advises the user 
of the number of unconverged layers.

\section{Uncertainties}
The code is equipped with different recipes to compute the 
uncertainties on each derived abundance. Several sources of 
error can affect the determination of chemical abundances, 
mainly the uncertainties due to the EW measurements and to the 
adopted log{\sl gf} (that are random errors from line to line) 
and those arising from the choice of the atmospheric parameters 
(that are random errors from star to star but systematic from line to line 
in a given star). These uncertainties are quantified by GALA while other 
sources of errors (as the choice of the abundance calculation code or 
the adoption of the grid of model atmospheres)
are neglected because considered {\sl external} errors.

\subsection{Abundances statistical errors}
The statistical uncertainty on the abundance of each element is computed 
by considering only the surviving lines after the rejection process 
(see Sect.~\ref{weed}). When the uncertainty on the EW is provided for each 
individual line (thus allowing to compute the abundance error for each transition), 
the mean abundance is computed 
by weighing the abundance of each line on its error, otherwise 
simple average and dispersion are computed. 
For those elements for which only one line is available, the error in abundance 
is obtained by varying the EW of 1$\sigma_{EW}$ (if the EW 
uncertainties are provided). Otherwise, the adopted value is zero.
As customary,  
the final statistical error on the abundance ratios is defined as $\sigma/\sqrt{N_{lines}}$.

\subsection{Uncertainties on the atmospheric parameters}
GALA estimates the internal error for each stellar parameter 
that has been derived from the spectroscopic analysis.
The uncertainties of $T_{\rm eff}$, $v_{\rm t}$ and log~g are estimated propagating 
the errors of the corresponding optimization parameter: 

$$\sigma_{\rm X_{\rm i}}=\frac{\sigma_{\rm C(X_{\rm i})}}{(\frac{\delta C(X_{\rm i})}{\delta X_{\rm i}})}$$
where $X_{\rm i}$ are $T_{\rm eff}$, log~g and $v_{\rm t}$, C($X_{\rm i}$) 
indicates the optimization parameters defined in Section~\ref{cxopt}, 
and $\sigma_{\rm C(X_{\rm i})}$ are the corresponding uncertainties.

The terms $\frac{\delta C(X_{\rm i})}{\delta X_{\rm i}}$ (which parametrize how the slopes and the iron difference 
vary with the appropriate parameters) are calculated numerically, by varying 
$X_{\rm i}$ locally around the final best value, assuming the step used in the optimization 
process and recomputing the corresponding 
$C(X_{\rm i})$.

The terms $C(X_{\rm i})$ are computed by applying a 
Jackknife bootstrapping technique \citep[see][]{lupton}.
The quoted quantities are recomputed by leaving out from the sample each time 
one different spectral line (thus, given a sample of N lines, each C(X) is computed 
N times by considering a sub-sample of N-1 lines). The uncertainty on the 
parameter X is $\sigma_{Jack}=\sqrt{N-1}\sigma_{sub}$, where $\sigma_{sub}$ 
is the standard deviation of the C(X) distribution derived from N sub-samples.
The $\sigma_{Jack}$ takes into account the uncertainty arising from the 
sample size and the line distribution, and this resampling method is especially useful 
to estimate the bias arising from the lines statistics. Note that the computation 
of the slopes is performed taking into account the effect of the EW and abundance 
uncertainties of each individual line.
Thus, the uncertainty in the atmospheric parameter $X_{\rm i}$ becomes
$$\sigma_{\rm X_{\rm i}}=\frac{\sigma^{Jack}_{\rm C(X_{\rm i})}}{(\frac{\delta C(X_{\rm i})}{\delta X_{\rm i}})}$$

It is worth to notice that these uncertainties 
represent the internal error in the derived parameters and are 
strongly dependent on the number of used lines and on
the distribution of the lines (weak and strong transitions for 
the estimate of $v_{\rm t}$ and low and high $\chi$ lines for $T_{\rm eff}$). 
Other factors that can affect the determination of the parameters 
(for instance, the threshold adopted in the EWs and in $\sigma_{EW}$) 
are not included in the error budgets and they can be considered as 
external errors. 
Finally, the error due to the adopted grid size could be considered 
as a systematic uncertainty (being the same for all the analysed stars) and 
eventually added in quadrature to the internal error estimated by GALA.

\subsection{Abundances uncertainties due to the atmospheric parameters}

The evaluation of the uncertainties arising from the atmospheric parameters is a more
complex task. Generally these errors are referred to as "systematic" 
uncertainties but this nomenclature is rather imprecise. In fact, the variation 
of a given parameter changes the abundance derived from the lines of the same 
element in a similar way (for instance, an increase of $T_{\rm eff}$ increases 
the abundance of all the iron lines). However, this error will be different from star to star, 
due to the different number of lines, strength and $\chi$ distributions, 
EWs quality and so on. Thus, the uncertainties from the atmospheric parameters 
should be considered as random errors, when we compare different stars 
(but they are systematic uncertainties from line to line).

Several recipes are proposed in the literature. The most common method is to 
re-compute the abundances changing each time one parameter only, and keeping 
fixed the other ones to their best estimates. Then, the corresponding variations 
in the abundances are added in 
quadrature. This approach is the most conservative, because it neglects the covariance terms 
arising from the interplay among the parameters (see Sect.~\ref{interplay}), 
providing only an upper limit for the total error budget. 

GALA follows the approach described by \citet{cayrel04} to naturally take into account 
the covariance terms. When the optimization process is ended, the analysis is 
repeated by altering the final $T_{\rm eff}$ by +$\sigma_{T_{\rm eff}}$ and -$\sigma_{T_{\rm eff}}$ 
(these uncertainties are calculated as described in Section 6.2), 
and re-optimizing the other parameters. The net variation of each chemical abundance with respect 
to the original value is assumed as final uncertainty due to the atmospheric 
parameters and including naturally the covariance terms.
Additionally, under request, also the abundance variations following the classical approach to vary 
one only parameter each time (keeping the other parameters fixed) are calculated,  
leaving the user free to use this information as preferred.

\subsection{Quality parameter for the final solution}

GALA provides also a check parameter, useful to judge the quality of the global solution
and to identify rapidly stars with unsatisfactory solutions.
For each model used during the optimization process a merit function $F_{merit}$ 
is defined as:
$$F_{merit}=\sqrt{{(\frac{S_{\chi}}{\sigma_{Jack}^{S_{\chi}}})}^{2} +  
{(\frac{S_{EWR}}{\sigma_{Jack}^{EWR}})}^{2}+
{(\frac{\Delta Fe}{\sigma_{Jack}^{\Delta Fe}})}^{2}},$$
taking into 
account the values of the optimization parameters and the corresponding uncertainties. 
In an ideal case, $F_{merit}$ is zero if the three optimization parameters 
are exactly zero. Generally, all the solutions with $F_{merit}\simeq$1 are 
valid and equally acceptable, while values of $F_{merit}>>$1 are suspect and 
point out that at least one of the parameters is not well constrained within the quoted 
uncertainty. Note that $F_{merit}$ provides only an indication if the solution is acceptable or not, 
but it does not specify which parameter is not well defined.


Summaryzing, when the full optimization process is completed, GALA will provide 
for each analysed element the (weighted) mean abundance, the dispersion and the 
number of used lines (which provide the statistical uncertainty), the net variation 
in abundance due to the new optimization with $T_{\rm eff}+\sigma_{T_{\rm eff}}$ and 
that with $T_{\rm eff}-\sigma_{T_{\rm eff}}$ 
(which provide the uncertainty owing to the choice of stellar parameters).
Also, for each atmospheric parameters the quoted internal uncertainties are computed.
Finally, the quality parameter $F_{merit}$ is provided to evaluate 
the goodness of the solution as whole.

\section{Dependence on SNR}

We performed a number of experiments to test the stability and reliability of 
the derived atmospheric parameters with GALA at different noise conditions. 
We performed two kind of 
experiments, described in the following: the first  based on a grid of 
synthetic spectra of abundances and atmospheric parameters 
known a priori, in order to estimate the reliability of the 
code as a function of the parameters and the signal-to-noise; 
the second group of tests is based on 
real spectra already analysed in literature.
In the following, the EWs were measured by means of DAOSPEC \citep{stetson} 
adopting a Gaussian profile for the line fitting.

\subsection{Synthetic spectra at different noise conditions}
\label{mcnoise}

We analysed with GALA a grid of synthetic spectra, computed to mimic 
the UVES@VLT high resolution spectra with the 580 Red Arm setup.
The grid of synthetic spectra includes SNR of 20, 30, 50, 100 per pixel for two different 
sets of atmospheric parameters: $T_{\rm eff}$=~4500 K, logg=1.5, $v_{\rm t}$=~2 km/s, [M/H]=~--1.0 dex 
to simulate a giant star, and $T_{\rm eff}$=~6000 K, logg=4.5, $v_{\rm t}$=~1 km/s, [M/H]=~--1.0 dex 
to simulate a dwarf star.
The spectra were computed with the following procedure:\\ 
(1)~for a given model atmosphere, two synthetic spectra were calculated with the 
SYNTHE code over the wavelength range covered by the two CCDs of the 580 UVES Red Arm grating 
and then convoluted with a Gaussian profile in order to mimic the formal UVES 
instrumental broadening;\\ 
(2)~the spectra were rebinned to a constant pixel-size ($\delta\lambda$=~0.0147 and 
0.0174 pixel/$\mathring{A}$ for the lower and upper chips respectively);\\ 
(3)~the synthetic spectra (normalized to unity) were multiplied with the efficiency curve 
computed by the FLAMES-UVES ESO Time Calculator in order to model the shape 
of the templates as realistically as possible;\\ 
(4)~Poissonian noise was injected in the spectra to simulate different noise conditions. 
Basically, the SNR varies along the spectrum, as a function of the efficiency (and thus of the 
wavelength). The noise was added in any spectrum according to the curve of SNR as 
a function of $\lambda$ provided by the FLAMES-UVES ESO Time Calculator.
For each SNR a sample of 200 synthetic spectra were generated.

Fig.~\ref{mc} summarizes the average values obtained for each MonteCarlo sample 
for each atmospheric parameter as a function of SNR; the errorbars indicate the 
dispersion around the mean. Results of the simulations of the giant star model atmosphere are 
shown in the upper panels of each window, while the lower panels summaryze the 
results for the dwarf star simulations.
Basically, the original parameters of the synthetic spectra (marked in Fig.~\ref{mc} 
as dashed horizontal lines) are recovered with small dispersions and 
without any significant bias. The major departure from the original values is observed 
in the microturbulent velocities (both dwarf and giant) at SNR=~20, because of
the loss of weak lines.

\begin{figure}[h]
\epsscale{1.2} 
\plotone{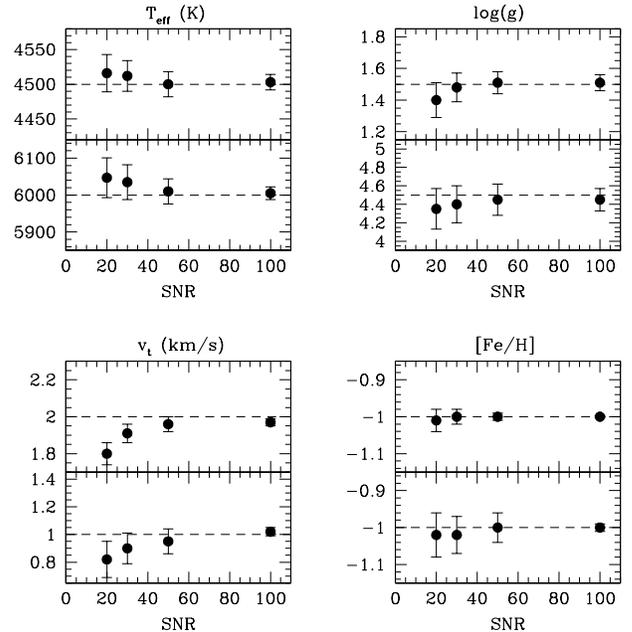}
\caption{Average values for the recovered atmospheric parameters as a function of SNR 
for the MonteCarlo samples described in Sect.~\ref{mcnoise}: upper panels of each window 
show the results for the giant star model, while lower panels show the results for 
the dwarf stars. Errorbars are the dispersion by the mean. Dashed lines are the 
original value for each parameter.}
\label{mc}

\end{figure}

\subsection{Real echelle spectra at different noise conditions}

The previous test provides an indication about the stability of the method 
against the SNR, starting from spectra their parameters are known 
{\sl a priori}. However, the employed noise model is a simplification 
because it does not take into account some effects that can 
also heavily affect the measurement of the EWs, such as the correlation of the 
noise among adjacent pixels, flat-fielding residuals, failures in the echelle orders 
merging, presence of spectral impurities. Also, the atomic data of the analysed lines are the 
same used in the computation of the synthetic spectra, thus excluding from the final 
line-to-line dispersion the random error due to the uncertainty on the atomic data.

In order to provide an additional 
test about the performance of GALA in conditions of different noise, we 
performed a simple experiment on the spectra acquired with UVES@FLAMES of the giants star 
NGC~1786-1501 in the Large Magellanic Cloud globular cluster NGC~1786 \citep[see][]{m09,m10}.
This is a sample of 8 spectra with the same exposure time ($\sim$45 min) obtained 
under the same seeing conditions was secured, with a typical SNR per pixel of 20 for each exposure. 
We used this dataset in order to obtain 8 spectra with different SNR ranging from $\sim$20 to $\sim$60 
depending on the number of exposures averaged: the spectrum with the lowest SNR is just one exposure, 
the spectrum with the largest SNR the average of all the 8 acquired exposures.
The derived parameters for each spectrum are showed in Fig.~\ref{lmc} as a function of SNR. 
The errorbars are derived by applying the Jackknife bootstrap technique for $T_{\rm eff}$, 
log~g and $v_{\rm t}$, 
while for [Fe/H] we used the dispersion by the mean as estimate of the error.

We note that the parameters are well constrained with small uncertainties 
for spectra with low SNR: this is not a numerical artifact of the code but it is due 
to the large number of transitions available in the UVES spectra, coupled with 
an accurate rejection of the outliers and the use of the uncertainties for each 
individual line in the slopes computations. 
Also, we note that major departures from the final parameters are again in the determination of $v_{\rm t}$ 
at SNR=~20: the derived low value of $v_{\rm t}$ is due to the fact that at low SNR several weak lines 
(useful to constrain the microturbulent velocity) are not well measured (then discarded by GALA) 
or not identified in the noise envelope by DAOSPEC. Note that the derived trend of $v_{\rm t}$ as a function of SNR 
shows the same behaviour found and discussed by \citet{m11rn}.

\begin{figure}[h]
\epsscale{1.2} 
\plotone{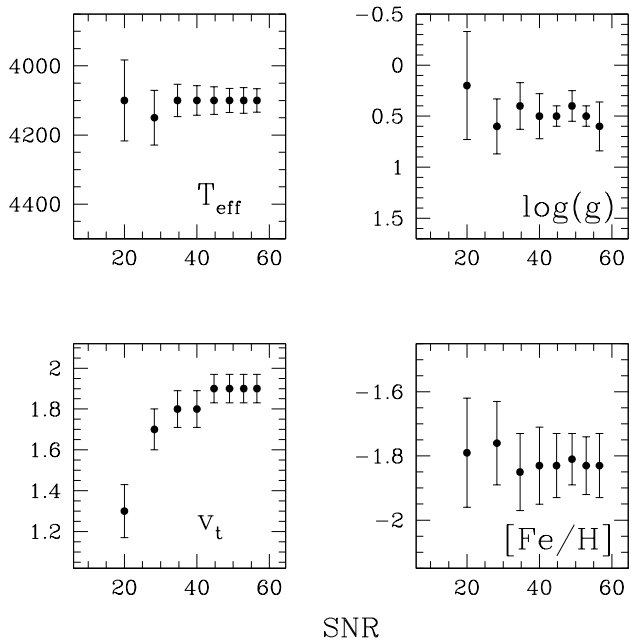}
\caption{Behaviour of the derived atmospheric parameters for the giant star 
NGC~1786-1501 as a function of SNR obtained by using different UVES co-added spectra. 
Errorbars are derived from the Jackknife bootstrap technique for $T_{\rm eff}$, log~g and $v_{\rm t}$, 
and as dispersion by the mean for [Fe~I/H].}
\label{lmc}
\end{figure}

\section{An efficient approach: Arcturus, Sun, HD~84937 and $\mu$Leonis}
\label{appro}

In this section we describe a convenient and robust method for performing abundance analysis with 
GALA, applied to the case of 4 stars (namely, the Sun, Arcturus, HD~84937 and $\mu$Leonis) 
of different metallicity and evolutionary stage and whose parameters 
are well established among the closest F-G-K stars. We retrieved high-resolution ($\sim$45000) spectra 
from the ESO\footnote{http://archive.eso.org/eso/eso$\_$archive$\_$main.html} 
(for Sun, Arcturus and HD~84937) and ELODIE\footnote{http://atlas.obs-hp.fr/elodie/}
 (for $\mu$Leonis) archives.

\subsection{Selection of the lines}
For each star we defined a suitable linelist of Fe~I and Fe~II transitions, starting from 
the most updated version of the Kurucz/Castelli lines dataset 
\footnote{http://wwwuser.oat.ts.astro.it/castelli/linelists.html}. 
We apply an iterative procedure to define the linelist. Assuming that the parameters 
of the targets are not known a priori, we performed a first analysis by using a preliminary linelist including 
only laboratory transitions with $\chi<$6 eV and log~(gf)$>$--5 dex.
Such a linelist is not checked against the spectral blendings arising from the adopted spectral resolution
and the atmospheric parameters and it is used only to perform a preliminary analysis. 
With the derived new parameters we define a new linelist for each star. 
The lines are selected by the inspection of synthetic spectra computed with 
the new parameters and convoluted with a Gaussian profile in order to reproduce the observed 
spectral resolution. At this step, only iron transitions predicted to be unblended are taken into account and used 
for the new analysis. \\

\subsection{EWs measurements} 
EWs are measured by using the code DAOSPEC which adopts 
a saturated Gaussian function to fit the line profile and an unique value for the full width half maximum (FWHM) 
for all the lines.
We start from the FWHM derived from the nominal spectral resolution of the spectra, leaving DAOSPEC free 
to re-adjust the value of FWHM according to the global residual of the fitting procedure.
The measurement of EWs is repeated by using the optmized FWHM value as a new input value, until convergence
is reached at a level of 0.1 pixel.
The formal error of the fit provided by DAOSPEC is used as 1$\sigma$ uncertainty on the EW measurement.\\

\subsection{Analysis with GALA}

The programme stars are analysed by employing all the three Working-Blocks, requiring 
a spectroscopic optimization of $T_{\rm eff}$, log~g, $v_{\rm t}$ and [M/H] and starting in all cases 
from the same set of guess parameters (namely $T_{\rm eff}$=~5000, logg=~2.5, [M/H]=~--1.0 dex 
and $v_{\rm t}$=~1.5 km/s). The optimization is performed by exploring the parameters space in small
steps of $\delta T_{\rm eff}$=~50 K, $\delta$log~g=~0.1 and $\delta v_{\rm t}$=~0.1 km/s; the metallicity is investigated 
by adopting the step of the ATLAS9 grids ($\delta$[M/H]=~0.5 dex). 

In the first run, we analysed the programme stars assuming the same 
configuration for the input parameters of GALA, in particular we included only lines 
with $\sigma_{EW}<$10\% and with EWR$>$--5.8 (corresponding to $\sim$10 m$\mathring{A}$
at 6000 $\mathring{A}$). After a first run of GALA, we refined the maximum allowed EWR, 
that depends mainly by the onset of the saturation along the curve of growth (and thus 
it is different for stars with different atmospheric parameters).
We adopted as maximum allowed value, EWR=~--4.65 for Arcturus and $\mu$Leonis,  
and --4.95 for the Sun and HD~84937. These values are chosen on the basis of the visual inspection
of the curve of growth, in order to exclude too strong lines, for which the Gaussian 
approximation can fail. After a first run of GALA, the linelist is refined by using the new parameters 
obtained by GALA as described above 
and the procedure repeated. Table 1 summaryzes the derived atmospheric parameters (with the 
corresponding Jackknife uncertainties) and the [Fe/H]~I and [Fe/H]~II abundance ratios, together with 
the two errorbars due to the atmospheric parameters and the internal error computed as $\sigma/\sqrt(N_{lines})$.

We compare our results with those available in literature (and listed in Table 1 as reference). 
We derive for the Sun $T_{\rm eff}$=~5800$\pm$64 K,  log~g=~4.50$\pm$0.18, 
$v_{\rm t}$=~1.20$\pm$0.13 km/s and [Fe/H]=--0.01$\pm$0.03 dex 
(where the errorbar is the sum in quadrature of the individual uncertainties listed in Table 1).
Our results for $T_{\rm eff}$ and log~g agree very well with those listed in the compilation 
of the NASA website\footnote{http://nssdc.gsfc.nasa.gov/planetary/factsheet/sunfact.html}.
Concerning the microtubulent velocities, 
values available in literature range from 0.8 km/s \citep{biemont} to 1.35 km/s \citep{steffen}.
A value of 1 km/s is typically adopted as representative for the Sun in several chemical analysis 
\citep[see][]{caffau}.

Our analysis of Arcturus provides $T_{\rm eff}$=~4300$\pm$60 K, log~g=1.60$\pm$0.06 
and $v_{\rm t}$=~1.50$\pm$0.06 km/s, 
with an iron abundance [Fe/H]=~-0.51$\pm$0.05. 
These results well match the recent analysis of Arcturus by
\citet{ramirez11} that provide an accurate determination of the atmospheric parameters and the chemical composition; 
in particular $T_{\rm eff}$ and log~g are derived in an independent way with respect to our approach, finding 
$T_{\rm eff}$=~4286$\pm$30 K (by fitting the observed spectral energy distribution), log~g=1.66$\pm$0.05 (through 
the trigonometric parallax), while $v_{\rm t}$ turns out to be 1.74 km/s (by using the same approach used in GALA).
The final iron abundance is [Fe/H]=~--0.52$\pm$0.04 dex. In both cases, the agreement with the literature 
values is good.
Finally, Fig.~\ref{galaplot} shows as example the graphical output of GALA for Arcturus.

\begin{figure}[h]
\epsscale{1.2} 
\plotone{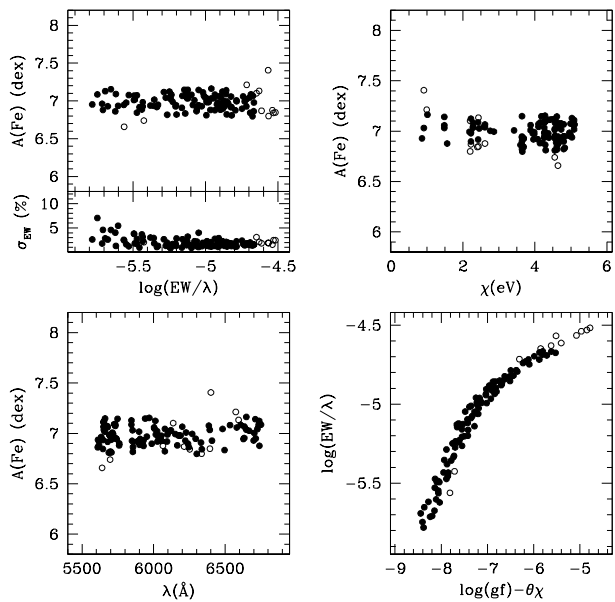}
\caption{Example of the graphical output of GALA for Arcturus (see Section \ref{appro}): 
black circles are the Fe~I lines used in the analysis and the empty circles show
rejected points.
}
\label{galaplot}
\end{figure}

For HD~84937 we derive $T_{\rm eff}$=~6150$\pm$56 K, log~g=~3.20$\pm$0.13, $v_{\rm t}$=~0.70$\pm$0.24 km/s
and [Fe/H]=~-2.28$\pm$0.06 dex, while for $\mu$Leonis we obtain $T_{\rm eff}$=~4500$\pm$81 K, log~g=2.40$\pm$0.26, 
$v_{\rm t}$=~1.40$\pm$0.07 km/s and [Fe/H]=~+0.37$\pm$0.06 dex.
For these two stars several determinations are available in literature and we decide to use as reference the average 
of the values listed in the classical compilation by \citet{cayrel01}, providing $T_{\rm eff}$=~6251$\pm$94 K, 
log~g=~3.97$\pm$0.18 and [Fe/H]=~--2.14$\pm$0.17 dex for HD~84937 and $T_{\rm eff}$=~4504$\pm$121 K, log~g=~2.33$\pm$0.27
and [Fe/H]=~+0.28$\pm$0.13 dex for $\mu$Leonis. 
Note that the value of the microturbulent velocities is omitted by \citet{cayrel01} because 
the listed authors in their compilation use different definitions for this parameter 
\citep[see][for a review of the different approaches]{m11rn} or assume a representative value. 
For HD~84937 the values of $v_{\rm t}$ range from 0.8 up to 1.7 km/s, while our value is slightly lower.
Also for $\mu$Leonis the range of values for  $v_{\rm t}$ is wide (from 1.2 up to 2.2 km/s) but our value 
agrees with this value.
Basically, the agreement with the values listed by \citet{cayrel01} is good, but for the gravity of HD~84937, for which 
we derive a lower value; this difference can be partially explained in light of the different $v_{\rm t}$.

\begin{deluxetable*}{cccccc}[htbp] 
\tablecolumns{6} 
\tablewidth{0pc}  
\tablecaption{Iron abundance (from neutral and singly ionized lines) and atmospheric parameters derived with GALA 
for Arcturus, $\rm \mu\rm{Leonis}$, the Sun and HD~84937. For the abundances, the first two errorbars 
are the uncertainties arising from the atmospheric parameters following the prescriptions by \citet{cayrel04}, 
while the last one is the internal error calculated as $\sigma/\sqrt(N_{lines})$. The uncertainties in the 
derived atmospheric parameters are computed with a Jackknife bootstrap technique.
In the lower part of the Table, the values available in literature are listed for comparison.}
\tablehead{ \colhead{Star}  & [Fe~I/H]  &[Fe~II/H] & $T_{\rm eff}$  & log~g & $v_{\rm t}$ \\
 & (dex)     &  (dex)   &  (K)       &       & (km/s) }
\startdata 
\hline
$\rm Sun$             &   --0.01$^{+0.02}_{-0.03}\pm0.009$   &  +0.01$^{+0.03}_{-0.07}\pm{0.003}$   &  5800$\pm$64   &  4.50$\pm$0.18  &  1.20$\pm$0.13 \\ 
$\rm Arcturus$        &   --0.51$^{+0.05}_{-0.04}\pm0.007$   & --0.52$^{+0.03}_{-0.05}\pm{0.015}$   &  4300$\pm$60   &  1.60$\pm$0.06  &  1.50$\pm$0.06 \\
HD~84937              &   --2.28$^{+0.06}_{-0.04}\pm0.007$   & --2.27$^{+0.05}_{-0.04}\pm{0.020}$   &  6150$\pm$56   &  3.20$\pm$0.13  &  0.70$\pm$0.24 \\ 
$\rm \mu\rm{Leonis}$  &    +0.37$^{+0.04}_{-0.06}\pm0.011$   &  +0.38$^{+0.04}_{-0.08}\pm{0.033}$   &  4500$\pm$81   &  2.40$\pm$0.26  &  1.40$\pm$0.07 \\
\hline 
\hline   
$\rm Sun$             &    0.00   & 0.00  &  5778   &  4.438  &  0.8 --- 1.35 \\
$\rm Arcturus$        &   --0.52$\pm$0.02   & --0.40$\pm$0.03   &  4286$\pm$30   &  1.66$\pm$0.05  &  1.74 \\
HD~84937              &   --2.14$\pm$0.17   & ---  &  6251$\pm$94   &  3.97$\pm$0.18  &  0.8 --- 1.7  \\
$\rm \mu\rm{Leonis}$  &    +0.28$\pm$0.13   & ---  &  4504$\pm$121   &  2.33$\pm$0.27  &  1.2 --- 2.2 \\
\hline 
\enddata 
\end{deluxetable*}

\subsection{Stability against the initial parameters}
\label{stabi}

A relevant feature of an automatic procedure to infer parameters and abundances 
is its stability against the input atmospheric parameters. 
In order to assess the effect of different first guess parameters, we analyse 
the spectrum of Arcturus by investigating a regular grid of input parameters
with $T_{\rm eff}$ ranging from 3800 to 4800~K (in steps of 100~K) and log~g from 1.0 to 2.2 (in steps of 0.1).
Fig.~\ref{grid} shows the grid of the input parameters in the $T_{\rm eff}$--log~g plane 
(empty points) with the position of the derived parameters (black points); the upper panel summarizes 
the results when GALA is used without the {\sl Refinement Working-Block}, while the lower panel 
shows the results obtained by employing also the refinement option. 
The recovered parameters cover a small range: in the first run the dispersion of the mean 
is of 46 K for $T_{\rm eff}$ and 0.09 for log~g, while these values drop to 25 K and 0.05 respectively,
when the {\sl Refinement Working-Block} is enabled.

\begin{figure}[h]
\epsscale{1.2} 
\plotone{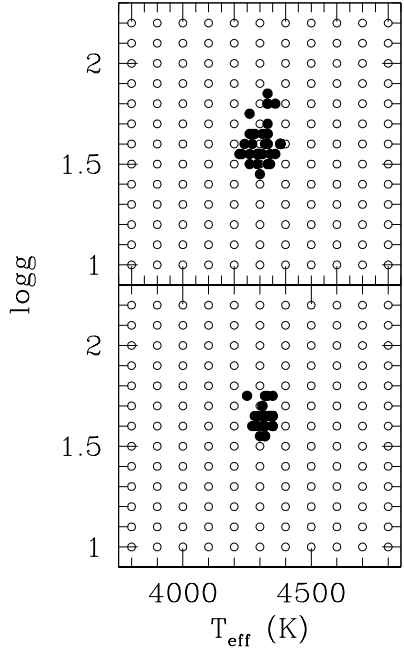}
\caption{Position of the final parameters for Arcturus (black points) in the $T_{\rm eff}$--logg plane, 
in comparison with the input parameters (empty circles), obtained by using GALA without (upper panel) and with 
(lower panel) the 
{\sl Refinement Working-Block} (upper and lower panel, respectively).}
\label{grid}
\end{figure}

\section{A test on globular clusters}

Globular clusters are ideal templates to check the capability 
of our procedure deriving reliable atmospherical parameters, because 
of the homogeneity (in terms of metallicity, age and distance) of their stellar content. Thus, the 
derived parameters for stars in a given globular cluster can be easy compared with theoretical 
isochrones in the $T_{\rm eff}$--log~g plane.

We apply the same procedure described in Sect.~\ref{appro} to analyse a set of high-resolution 
spectra for stars in the globular cluster NGC~6752, ranging from the Turn-Off up to the 
bright portion of the Red Giant Branch. The spectra have been retrieved by the ESO 
archive\footnote{http://archive.eso.org/cms/eso-data.html} 
and reduced with the standard ESO pipeline\footnote{http://www.eso.org/sci//software/pipelines/}.  
They are from different observing programmes and with different SNR, including 
giant stars crossing the Red Giant Branch Bump region observed with UVES@VLT (slit mode) within the ESO Large Program 
65.L-0165 with very high ($>$200) SNR, the stars in the bright portion of the 
Red Giant Branch observed with UVES-FLAMES@VLT (fiber mode) within the Galactic globular clusters survey 
presented by \citet{carretta} and the dwarf/subgiant stars observed with UVES@VLT (slit mode) 
within the ESO Large Program 165.L-0263. %

The main panel of Fig.~\ref{6752} shows the position of the final parameters derived with GALA in the 
$T_{\rm eff}$--log~g plane. 
Also, two theoretical isochrones with an age of 12 Gyr and a metallicity of Z=~0.0006 
\citep[assuming an $\alpha$-enhanced chemical mixture) are shown as reference (grey curve is from BaSTI database by
\citet{pietrinferni} and black curve from Padua database by][]{girardi}.
In the lower panel, we show the behaviour of the [Fe/H] ratio as a function of $T_{eff}$. 
No significant trend is found, while the star-to-star scatter increases with increasing $T_{eff}$ 
because of the lower SNR.

\begin{figure}[h]
\epsscale{1.2} 
\plotone{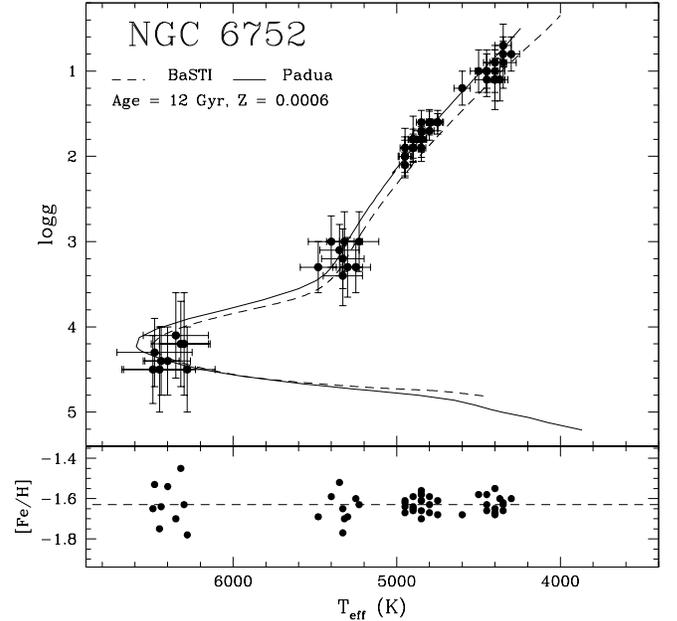}
\caption{Main panel: position in the $T_{\rm eff}$--log~g plane of the stars in the globular cluster NGC~6752 analysed with GALA . 
We plotted as references two isochrones computed with an age of 12 Gyr and a metallicity of Z=~0.0006 
(assuming an $\alpha$-enhancement chemical mixture), from the BaSTI \citep[][dotted curve]{pietrinferni} 
and Padua \citep[][solid curve]{girardi} database. 
Lower panel: behaviour of the [Fe/H] ratio as a function of $T_{\rm eff}$.}
\label{6752}
\end{figure}

The parameters derived with GALA well reproduce the behaviour predicted by the theoretical models for a old simple 
stellar population with the same metallicity of the cluster, confirming the physical reliability of the 
final solution.
Also, we note that the errorbars, both in $T_{\rm eff}$ and log~g, change according to the quality of the spectra, 
ranging from $\sim$30 K and $\sim$0.15 for the giants with the highest SNR up to $\sim$200 K and $\sim$0.5 
for the dwarf stars with the lower SNR.

The entire sample of 52 stars provides an average iron abundance of [Fe/H]I=--1.63 dex ($\sigma$=~0.06 dex). 
This value is consistent with the previous estimates available in literature that point out  
an iron content ranging from [Fe/H]=--1.62 \citep{grundhal} up to [Fe/H]=--1.42 dex \citep{gratton01}.
In this comparison we cannot take into account the different adopted solar values.\\
The stars in common with \citet{carretta} show a reasonable agreement in the atmospheric parameters, 
with $T_{eff}^{GALA}$-$T_{eff}^{Carretta}$=+49 K ($\sigma$=~42 K), 
$logg^{GALA}$-$logg^{Carretta}$=--0.26 ($\sigma$=~0.07) and 
$[Fe/H]^{GALA}$-$[Fe/H]^{Carretta}$=--0.08 dex ($\sigma$=~0.05 dex).\\
Also for the stars in common with \citet{yong} the agreement is excellent (also because these spectra 
have very high SNR and a large wavelength coverage, thus allowing the measurements of a large number 
of Fe~I and Fe~II lines):  
$T_{eff}^{GALA}$-$T_{eff}^{Yong}$=--1 K ($\sigma$=~32 K), 
$logg^{GALA}$-$logg^{Yong}$=--0.39 ($\sigma$=~0.11) and 
$[Fe/H]^{GALA}$-$[Fe/H]^{Yong}$=+0.00 dex ($\sigma$=~0.06 dex).\\
The comparison with the analysis of the Turn-Off and Sub-Giant Branch stars by \citet{gratton01} 
is not trivial, because they derived the atmospheric parameters from median spectra for the two groups of stars. 
The temperatures have been derived by fitting the wings of the $H_{\alpha}$, while the gravities 
have been obtained from the position of the stars in the Color-Magnitude Diagram. 
When we compared the parameters by \citet{gratton01} with the average values obtained by our analysis 
of individual stars, the agreement is not perfect but consistent within the uncertainties: 
we find $<T_{eff}^{GALA}>$-$T_{eff}^{Gratton}$=+164 K, $<logg^{GALA}>$-$logg^{Gratton}$=+0.06 
and $[Fe/H]^{GALA}$-$[Fe/H]^{Gratton}$=--0.19 dex ($\sigma$=~0.15 dex) for the Turn-Off stars, 
and $<T_{eff}^{GALA}>$-$T_{eff}^{Gratton}$=-15 K, $<logg^{GALA}>$-$logg^{Gratton}$=--0.36 
and $[Fe/H]^{GALA}$-$[Fe/H]^{Gratton}$=--0.23 dex ($\sigma$=~0.11 dex) for the Sub-Giant stars.



\section{Summary}

In this paper we have presented a new, automatic tool to perform accurate analysis of stellar spectra.
GALA is designed to perform automatically the search for the best atmospheric parameters 
$T_{eff}$, log~g, $v_t$ and the overall metallicity [M/H] for moderate and high resolution stellar 
absorption spectra, by using the EWs of metallic lines. 
Also, GALA provides the abundance of each individual line for which the user provides 
the EW, as well as the average abundance for each atomic species.

The source code of GALA is freely available at the website 
 {\sl http://www.cosmic-lab.eu/Cosmic-Lab/Products.html}, together 
 with the user manual (including the information about installation, 
 configuration of the input files and how to obtain and properly use 
 the model atmospheres) and an example of the input files as reference.
The main advantages of the code are the capability: 
\begin{enumerate}
\item to optimize all the parameters or only part of them. 
The code is versatile in order to perform different kind of analysis 
(full or partial spectroscopic analysis, experiments about the guess parameters...) and adopting 
different recipes to derive log~g;
\item to perform a careful rejection of the outliers according to the line strength, the EW quality 
and the line distribution in the A(Fe)--$\chi$ and A(Fe)--EWR planes;
\item  to estimate for each individual star the internal errors for
{\sl (a)}~the optimized parameters by adopting the Jackknife bootstrapping technique, 
and {\sl (b)}~the derived uncertainties due to the choice of atmospheric parameters,
following both the prescriptions 
by \citet{cayrel04} and the classical method of altering one parameter at a time.
\end{enumerate}

We have performed an extensive set of test with both synthetic and observed spectra, in order 
to assess the performances of the code. 

\begin{enumerate}
\item

Experiments with synthetic spectra 
(whose atmospheric parameters are known a priori) with the injection of Poission noise 
to simulate different noise conditions show a high stability of the code to recover the 
atmospheric parameters, without significant bias. The major departure from the original values 
is found in the microturbulent velocity of low ($\sim$20) SNR spectra, due to the systematic loss of 
weak lines.

\item
A set of FLAMES-UVES spectra of the LMC giant star NGC~1786-1501 observed with different 
signal-to-noise (from $\sim$20 up to $\sim$60) has been analysed with GALA, confirming that 
our procedure well constrains 
the parameters also in case of low spectral quality. Also in this case, we found that the 
largest departures are for the microturbulent velocity in the spectra with low SNR, because 
the weak lines are not well measured or not detectable in the noise envelope, leading to 
an underestimate of this parameter.

\item
We analysed 4 stars (namely, the Sun, Arcturus, HD~84937 and $\mu$Leonis) 
of different metallicity and evolutionary stage and whose parameters 
are well established among the closest F-G-K stars. We described an efficient method 
(including the line selection, the measurement of the EWs and the chemical analysis) 
to best exploit the capabilities of GALA. Our results for these stars (both for atmospheric 
parameters and [Fe/H] ratio) well agree with those available in literature.

\item 
Finally, we analysed a sample of 52 stars of the Galactic globular cluster NGC~6752, 
in different evolutionary stages. The derived $T_{eff}$ and log~g well follow the predictions 
of theoretical isochrones with the appropriate age and chemical composition for this cluster. 
Also, the derived iron content nicely agrees with the previous estimates of other works. 
\end{enumerate}

The code permits to obtain chemical abundances and atmospheric parameters for large stellar 
samples in a very short time, thus making GALA an useful tool in the epoch of the multi-object 
spectrographs and large surveys. 
Because of its nature of open source code, GALA will be implemented in the next releases 
according to the feedback with the users. In particular, we plan to include in the code 
new grids of ODFs and models that will be publicly released in the future. Also, GALA 
will be constantly updated in order to include variations and changes in ATLAS9 and MARCS 
models, as well as in the ATLAS9 code.

\acknowledgements  

The authors warmly thank Fiorella Castelli, Thomas Masseron, Piercarlo Bonifacio and Andrea Negri for 
useful comments, discussions and suggestions.
We thanks the anonymous referee for a careful reading of the paper and 
helpful comments.
This research is part of the project COSMIC-LAB funded by the European Research Council 
(under contract ERC-2010-AdG-267675).
\\

{\it Und wenn dich das Irdische verga\ss, zu der stillen Erde sag: Ich rinne. 
Zu dem raschen Wasser sprich: Ich bin. (R. M. Rilke)}


\newpage

\begin{thebibliography}{}

\bibitem[Allende Prieto et al.(2008)]{allende08}
Allende Prieto, C., et al., 2008, AN, 329, 1018
\bibitem[Barden et al.(2010)]{barden10}
 Barden, S. C., et al., 2010, SPIE, 7735, 8
 \bibitem[Biemont et al.(1981)]{biemont}
Biemont, E., Grevesse, N., Hannaford, P., \& Lowe, R. M., 1981, ApJ, 248, 867
\bibitem[Bonifacio \& Caffau(2003)]{boni03}
Bonifacio, P., \& Caffau, E., 2003, A\&A, 399, 1183
\bibitem[Caffau et al.(2011)]{caffau}
Caffau, E., Ludwig, H.-G., Steffen, M., Freytag, B., \& Bonifacio, P., 2011, SoPh, 268, 251
\bibitem[Carretta et al.(2009)]{carretta}
Carretta, E., Bragaglia, A., Gratton, R., \& Lucatello, S., 2009, A\&A, 
505, 139
\bibitem[Castelli(1988)]{castelli88}
Castelli, F., 1988, Osservatorio Astronomico di Trieste, 1164
\bibitem[Castelli \& Kurucz(2004)]{castelli04}
Castelli, F., \& Kurucz, R. L., 2004, arXiv:astro-ph/0405087v1
\bibitem[Castelli(2005a)]{castelli_w}
Castelli, F., 2005, MSAIS, 8, 44
\bibitem[Castelli(2005b)]{castelli_12}
Castelli, F., 2005, MSAIS, 8, 25
\bibitem[Cayrel de Strobel, Soubiran \& Ralite(2001)]{cayrel01}
Cayrel, de Strobel, G., Soubiran, C., \& Ralite, N., 2001, A\&A, 373, 159
\bibitem[Cayrel et al.(2004)]{cayrel04}
Cayrel, R., et al., 2004, A\&A, 416, 1117
\bibitem[Castelli \& Kurucz(2004)]{kur04}
Castelli, F., \& Kurucz, R. L., 2004, astro.ph.5087
\bibitem[Cirasuolo et al.(2011)]{cirasuolo11}
Cirasuolo, M., Afonso, J., Bender, R., Bonifacio, P., Evans, C., Kaper, L., 
Oliva, E., \& Vanzi, L., 2011, Messenger, 145, 11
\bibitem[de Jong(2011)]{dejong11}
de Jong, R., 2011, Messenger, 145, 14
\bibitem[Edvardsson(1988)]{edv88}
Edvardsson, B., 1988, IAUS, 132, 387
\bibitem[Fitzpatrick \& Sneden(1987)]{sneden}
Fitzpatrick, M. J. \& Sneden, C., 1987, BAAS, 19, 1129
\bibitem[Fontenla, Avrett \& Loeser(1993)]{fontenla}
Fontenla, J. M., Avrett, E. H., \& Loeser, R., 1993, ApJ, 406, 319
\bibitem[Freeman et al.(2013)]{freeman13}
Freeman, K., et al., 2013, MNRAS, 428, 3660
\bibitem[Gilmore et al.(2012)]{gilmore12}
Gilmore, G. et al., 2012, Messenger, 147, 25
\bibitem[Girardi et al.(2000)]{girardi}
Girardi, L., Bressan, A., Bertelli, G., \& Chiosi, C., 2000, A\&AS, 141, 371
\bibitem[Gratton et al.(2001)]{gratton01}
Gratton, R. et al., 2001, A\&A, 369, 87
\bibitem[Grundahl et al.(2002)]{grundhal}
Grundahl, F., Briley, M., Nissen, P. E., \& Feltzing, S., 2002, A\&A, 385, L14
\bibitem[Gustafsson et al.(2008)]{gustaf98}
Gustafsson, B., Edvardsson, B., Eriksson, K., Jorgensen, U. G., Nordlund, A., \& 
Plez, B., 2008, A\&A, 486, 951
\bibitem[Kunder et al.(2012)]{kunder}
Kunder, A. et al., 2012, AJ, 143, 57
\bibitem[Kurucz(2005)]{kurucz05}
Kurucz, R. L., 2005, MSAIS, 8, 14
\bibitem[Lupton(1993)]{lupton}
Lupton, R., 1993, in Statistics in Theory an Pratice (Princeton, NJ:
Princeton, Univ. Press)
\bibitem[Masseron(2006)]{masse}
Masseron, T. 2006, PhD thesis, Observatoire de Paris, France
\bibitem[Meszaros et al.(2012)]{meszaros}
Meszaros Sz. et al. ,2012, AJ, 144, 120
\bibitem[Mucciarelli et al.(2009)]{m09}
Mucciarelli, A., Origlia, L., Ferraro, F. R., \& Pancino, E., 2009, ApJ, 695L, 134
\bibitem[Mucciarelli et al.(2010)]{m10}
Mucciarelli, A., Origlia, L., \& Ferraro, F. R., 2010, ApJ, 717, 277
\bibitem[Mucciarelli(2011)]{m11rn}
Mucciarelli, A., 2011, A\&A, 528, 44
\bibitem[Pietrinferni et al.(2006)]{pietrinferni}
Pietrinferni, A., Cassisi, S., Salaris, M., \& Castelli, F.\ 2006, ApJ, 642, 797 
\bibitem[Posbic et al.(2012)]{posbic}
Posbic, H., Katz, D., Caffau, E., Bonifacio, P., Gomez, A., Sbordone, L.,
\& Arenou, F., 2012, arXiv1209.0407
\bibitem[Press et al.(1992)]{press}
Press.,W. H., Teukolsky, A. A., Vetterling, W. T., \& Flannery, B. P., 
Numerical Recipes, 2nd edn. (Cambridge: Cambridge Univ. Press)
\bibitem[Ramirez et al.(2001)]{ramirez01}
Ramirez, S. V., Cohen, J. G., Buss, J., \& Briley, M. M., 2001, AJ, 122, 1429
\bibitem[Ramirez \& Allende Prieto(2011)]{ramirez11}
Ramirez, I., \& Allende Prieto, C., 2011, ApJ, 743, 135
\bibitem[Recio-Blanco, Bijaoui \& de Laverny(2006)]{recio06}
Recio-Blanco, A, Bijaoui, A., \& de Laverny, P., 2006, MNRAS, 370, 141
\bibitem[Sbordone et al.(2004)]{sbordone04}
Sbordone, L., Bonifacio, P., Castelli, F., \& Kurucz, R. L., 2004, MeMSai, 5, 93
\bibitem[Sbordone et al.(2010)]{sbordone10}
Sbordone, L., Bonifacio, P., Caffau, E. \& Ludwig, H.G., 2010, arXiv1009.5210
\bibitem[Sousa et al.(2007)]{sousa07}
Sousa, S. G., Santos, N. C., Israelian, G., Mayor, M. \& Monteiro, M. J. P. F. G., 2007, A\&A, 469, 783
\bibitem[Steinmetz et al.(2006)]{steinmetz}
Steinmetz, M., et al., 2006, AJ, 132, 1645
\bibitem[Steffen, Ludwig \& Caffau(2009)]{steffen}
Steffen, M., Ludwig, H.-G., \& Caffau, E., 2009, MSAIS, 80, 731
\bibitem[Stetson \& Pancino(2008)]{stetson}
Stetson, P. B., \& Pancino, E., 2008, PASP, 120, 1332
\bibitem[Valenti \& Piskunov(1996)]{valenti}
Valenti, J. A., \& Piskunov, N., 1996, A\&AS, 118, 595
\bibitem[Yong et al.(2005)]{yong}
Yong, D., Grundahl, F., Nissen, P. E., Jensen, H. R., \& Lambert, D. L., 2005, A\&A, 438, 875
\end{thebibliography}
\end{document}